\documentclass[12pt]{iopart}
\usepackage{geometry}
\usepackage{graphicx}         
\usepackage{ragged2e}
\usepackage{cite}       
\usepackage{amssymb}
\usepackage[caption=false]{subfig}
\usepackage{color}
\begin{document}


\title[Radiation friction induced enhancement of laser-driven longitudinal fields]{Theory and simulations of radiation friction induced enhancement of laser-driven longitudinal fields}
\author{E G Gelfer$^{1,2}$, A M Fedotov$^2$, S Weber$^1$}
\address{$^1$ELI Beamlines, Institute of Physics of the ASCR, v.v.i., Dolni Brezany, Czech Republic}
\address{$^2$National Research Nuclear University ``MEPhI'' (Moscow Engineering Physics Institute), 115409, Kashirskoe sh, 31, Moscow, Russia}
\ead{egelfer@gmail.com}

\begin{abstract}
We consider the generation of a quasistatic longitudinal electric field by intense laser pulses propagating in a transparent plasma with radiation friction taken into account. For both circular and linear polarization of the driving pulse we develop a 1D analytical model of the process, which is valid in a wide range of laser and plasma parameters. We define the parameter region where radiation friction results in an essential enhancement of the longitudinal field. The amplitude and the period of the generated longitudinal wave are estimated and optimized. Our theoretical predictions are confirmed by 1D and 2D PIC simulations. We also demonstrate numerically that radiation friction should substantially enhance the longitudinal field generated in a plasma by a 10 PW laser such as ELI Beamlines.
\end{abstract}



\submitto{\PPCF}
\maketitle

\section{Introduction}

The modern laser systems have recently reached the multi-Petawatt power level \cite{laserreview, chu,sung,zheng}. These facilities already provide extremely strong electromagnetic fields $\simeq 10^9$V/cm \cite{hercules}, but even more powerful 10 PW lasers are now under construction in Czech Republic (ELI Beamlines \cite{ELI}), Romania (ELI NP \cite{ELINP}) and France (Apollon \cite{Apollon}). However, since laser fields oscillate with very high frequency, their application to charged particle acceleration is not straightforward. By separating charges, a laser pulse propagating in plasma creates a wake (longitudinal) wave. As suggested many years ago \cite{tajima} and demonstrated in a recent experiment \cite{bellaacc}, this wake wave can efficiently accelerate electrons up to several GeVs. 

Several models were suggested also for ion acceleration in various laser-plasma setups \cite{light_sail,hole_boring}, see the reviews \cite{ion_acc_rev} and the overview of the recent experimental achievements in \cite{ion_acc_exp}. In all such schemes the laser pulse pushes plasma electrons forward via the ponderomotive force, thus leading to charge separation and creation of a quasistatic longitudinal electric field \cite{micha2012} capable for particle acceleration. However, as recently demonstrated \cite{arxiv}, besides the usual ponderomotive mechanism (PM) of charge separation in plasma, there also exists a competing radiation friction mechanism (RFM), which for strong and long laser pulses propagating in low density plasmas should dominate over PM. 

To clarify its origin, consider an electron in a circularly polarized plane wave. If radiation friction (RF) is neglected, the electron circles with the field frequency, so that its velocity remains perpendicular to the electric field and parallel to the magnetic field \cite{akhiezer_polovin}, with no force acting in the direction of pulse propagation. RF in transverse direction changes the angle between the electron velocity and the magnetic field \cite{arxiv, RFpapers}, resulting in the longitudinal Lorenz force $[\mathbf{v}\times\mathbf{B}]$\footnote{We use CGS-like units but with speed of light $c=1$.}, which accelerates the electron in the direction of pulse propagation, thus separating charges in plasma. 

The role of RF in laser-plasma interaction was previously a subject of many investigations \cite{zhidkov2002,tamburini2010,tamburini2011,brady2012,nakamura2012,bashinov2013,stark2016}, but mostly for overcritical plasmas. However, if plasma is so dense that it is opaque, then electrons never get deep inside the pulse, where the field is so strong that RF affects the electron transverse motion. As for the overcritical but relativistically transparent case, in a high density plasma RF results in a fast depletion of the laser pulse. Therefore we consider plasma density not exceeding the critical value $n_c=m\omega^2/4\pi e^2$, where $\omega$ is the laser frequency, $m$ and $e$ are the electron mass and charge.

\begin{figure}[t!]
\centering
\includegraphics[width =0.6\linewidth]{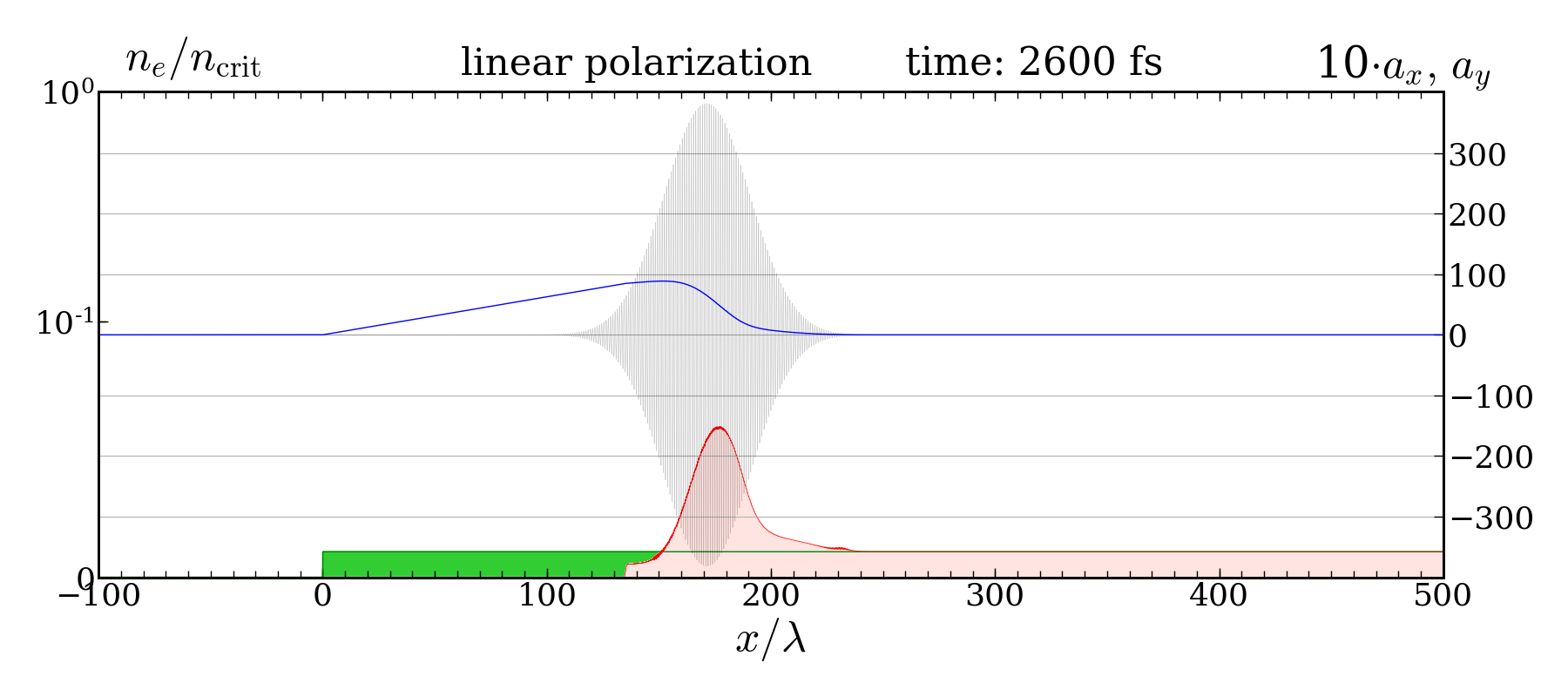}
\caption{\label{fig_spike}(Color online). 1D PIC simulation of laser pulse propagation through a plasma with  RF. Red and green filled areas -- electron and ion densities in units of the critical density $n_c$(quantified on the left); grey line -- $y$ component of the dimensionless transverse electric field $a_y=eE_y/m\omega$ (right axis); blue curve -- longitudinal component of the dimensionless electric field $a_x=eE_x/m\omega$, multiplied by the factor of $10$ (right axis). Laser intensity $I=2\cdot10^{23}$ W/cm${}^2$, pulse duration (FWHM) $t_{pulse}=150$ fs, linear polarization, initial plasma density $n=0.01 n_c$, immobile ions. The snapshot is given at time moment $t=2600$fs.}
\end{figure}

In such a regime charge separation and generation of the quasistatic longitudinal electric field by an intense laser pulse in a plasma with immobile\footnote{For discussion of the validity of this approximation see Section~\ref{sec_ca} below.} ions are perfectly described by the following 1D model \cite{arxiv}. A pulse entering a plasma accelerates the electrons forward via RFM and/or PM, piling them up into a moving density spike. Thereupon the longitudinal electrostatic field of the naked immobile ions is gradually growing, and eventually starts decelerating the electrons. Finally a breakdown occurs, when a part of electrons from the spike penetrates through the pulse backward partially screening the ion field, see Figure~\ref{fig_spike}. The amplitude and the period of the thus generated longitudinal field are defined by the position of the spike at the moment of breakdown, and to find it we consider the motion of the leftmost particle in the spike.  

The model developed in \cite{arxiv} provides estimations for the amplitude and the period of the longitudinal field generated by a circularly polarized laser pulse via both PM and RFM under the additional assumption of ultrarelativistic electron longitudinal motion, that are in a rather good agreement with PIC simulations. Here we discuss this model in more detail and generalize it to the cases of arbitrary polarization of the laser pulse and arbitrary electron longitudinal velocities, thus extending its validity to a wider range of laser parameters. We formulate and study the limits of applicability of the model (in particular establish the conditions required to consider the ions immobile), derive the estimations for various regimes of longitudinal field generation, and identify an optimal regime with generation of the highest attainable longitudinal field. Furthermore, by 2D PIC simulations we demonstrate that our 1D model is valid for wide pulses (of comparable length and width), and that RFM can dominate over PM also for tightly focused pulses with the parameters of the upcoming facilities such as ELI Beamlines \cite{ELI}.

\section{1D model for arbitrary laser polarization}

Let us start with a generalization of the model suggested in \cite{arxiv} to describe longitudinal electric field generation by a strong laser pulse propagating in plasma from pure circular to arbitrary polarization of the pulse.  The 4-dimensional form of the equation of motion for the leftmost particle in the electron spike (see Figure~\ref{fig_spike}) reads:
\begin{equation}\label{eqll}\fl
\frac{du^i}{ds}=\frac{e}{m}F^{ij}u_j+\frac{2e^3}{3m^2}\frac{\partial F^{ij}}{\partial x^l}u_ju^l-\frac{2e^4}{3m^3}F^{il}F_{jl}u^j+\frac{2e^4}{3m^3}(F_{jl}u^l)(F^{jm}u_m)u^i+\mathcal{F}^i.
\end{equation}
Here $F_{ij}=\partial_i A_j-\partial_j A_i$ is the laser field strength tensor, $\mathcal{F}^i=-4\pi e^2 n x\gamma/m\{v_x,1,0,0\}$ is the electrostatic 4-force from the naked ions; $s$, $\gamma$, $\mathbf{v}$ and $u^i=\{\gamma,\gamma\mathbf{v}\}$ are the proper time, gamma factor, 3- and 4-velocities of the electron, and we use the Landau-Lifshitz form \cite{LL} for RF force. Assume that the laser field is of the form of a plane wave, $A_i=A_i(\varphi)$, where $\varphi=k^ix_i$ is the phase and $k^i=\omega\{1,1,0,0\}$ is the wave 4-vector. By  contracting (\ref{eqll}) with $k_i$ and taking into account $F_{ij}=k_iA_j'(\varphi)-k_jA_i'(\varphi)$ and $k_ik^i=k_iA^i=k_iF^{ij}=0$, we arrive at
\begin{equation}\label {eq_ku}
\omega\frac{du_-}{ds}=\frac{2e^4\omega^3}{3m^3}\left(\frac{dA_i}{d\varphi}\frac{dA^i}{d\varphi}\right)u_-^3 + k_i\mathcal{F}^i,
\end{equation}
where $u_-=k^iu_i/\omega=\gamma-u_x$. From $u_iu^i=1$ and $dt=\gamma\,ds=(\gamma/\omega u_-)\,d\varphi$ one can easily derive the formulas 
$$\frac{du_-}{ds}=u_-\left(\frac{\omega}{2\gamma}\frac{du_\perp^2}{d\varphi}-\frac{du_x}{dt}\right),\quad u_-^2=(1+u_\perp^2)\frac{(1-v_x)}{(1+v_x)},$$
using which follows
\begin{equation}\label{dux}
\frac{du_x}{dt}=\frac{\omega}{2\gamma}\frac{d(u_\perp^2)}{d\varphi}-\frac{2e^4\omega^2}{3m^3}\left(\frac{dA^i}{d\varphi}\right)^2(1+u_\perp^2)\left(\frac{1-v_x}{1+v_x}\right)-\frac{4\pi e^2n}{m}x.
\end{equation}
An advantage of the above derivation is its generality with respect to polarization of the driving laser field, on which no assumptions have been made yet. We now consider the most important cases of circular (CP) and linear (LP) polarizations.

\subsection{Circular polarization}

If the laser pulse is circularly polarized $\mathbf{A}(\varphi)=(m/e)a_0(\varphi)\{\cos\varphi,\sin\varphi,0\}$  then $(dA^i/d\varphi)^2=-m^2a_0^2/e^2$, where $a_0(\varphi)$ is a slowly varying dimensionless pulse envelope. Denoting the dimensionless variables $\tau=\omega t$, $\xi=\omega x$ and parameters $\tilde{n}=n/n_c$, $\mu=2\omega r_e/3c\simeq 1.18\cdot 10^{-8}$, where $r_e=e^2/mc^2$ is the classical electron radius, we can cast (\ref{dux}) into the form
\begin{equation}\label{dux_c}
\frac{du_x}{dt}=\frac1{2\gamma}\frac{d(u_\perp^2)}{d\varphi}+\mu a_0^2(1+u_\perp^2)\frac{1-v_x}{1+v_x}-\tilde{n}\xi.
\end{equation}
Equations~(\ref{dux}) and (\ref{dux_c}) are the exact implications of (\ref{eqll}) and hence up to this moment no approximations except for the Landau-Lifshitz approximation and the assumption of a dilute plasma have been made. But if we further assume that damping due to RF is small (see Sec.~\ref{sec_ca} for discussion) then $u_\perp\approx a_0$ and, neglecting also $1$ against $a_0^2$, we can further reduce (\ref{dux_c}) to the form 
\begin{equation}\label{eqcp}
\frac{du_x}{d\tau}=\frac1{2\gamma}\frac{d(a_0^2)}{d\varphi}+\mu a_0^4\frac{1-v_x}{1+v_x}-\tilde{n}\xi.
\end{equation}
This is precisely the 1D model suggested in \cite{arxiv}. As mentioned in the Introduction, there are two mechanisms of electron longitudinal acceleration and charge separation in a plasma: PM [corresponding to the first term in the RHS of (\ref{eqcp})], and RFM [corresponding to the second term proportional to $\mu$]. In order to further simplify (\ref{eqcp}), consider two opposite limiting cases of the electron longitudinal motion: nonrelativistic ($u_x\ll a_0$) and ultrarelativistic ($u_x\gg a_0$). 

\subsubsection{Nonrelativistic longitudinal electron motion.}

In the nonrelativistic case $v_x\ll1$ we have $\gamma\approx a_0$, $u_x\approx v_x a_0$, $\varphi\approx \tau$, and (\ref{eqcp}) is the same as for a damped harmonic oscillator with slowly varying parameters and shifted equilibrium position:
\begin{equation}\label{eqnr}\fl
\ddot{\xi}+2\varkappa\dot{\xi}+\Omega^2\xi=f,\quad \varkappa(\tau)=\mu a_0^3+\frac{\dot{a}_0}{2a_0},\quad \Omega(\tau)=\sqrt{\frac{\tilde{n}}{a_0}},\quad f(\tau)=\mu a_0^3+\frac{\dot{a}_0}{a_0},
\end{equation}
where the dots abbreviate the derivatives with respect to $\tau$. In dilute plasmas under consideration here this effective oscillator is always over-damped ($\varkappa\gtrsim\Omega$) and very rapidly (within $\tau\lesssim \varkappa/\Omega^2$) approaches its equilibrium 
\begin{equation}\label{xi}
\xi_0(\tau)\simeq \frac{f(\tau)}{\Omega^2(\tau)}=\frac{\mu a_0^4(\tau)+\dot{a}_0(\tau)}{\tilde{n}}. 
\end{equation}
Assuming that 
\begin{equation}\label{nrcond}
T\gg \frac{\varkappa}{\Omega^2}\simeq \max\left\{\frac{\mu a_0^4}{\tilde{n}},\frac{a_0}{\tilde{n}T}\right\}, 
\end{equation}
where $T=\omega t_{pulse}$ is the dimensionless pulse duration, we can assume that the oscillator is settled at the equilibrium (by occasion, exactly the same condition validates the nonrelativistic regime which is under assumption).

Hence the amplitude of longitudinal charge separation field $a_\parallel^{(nr)}=\tilde{n}\xi^{max}$ can be estimated as 
\begin{equation}\label{anr}
a_\parallel^{(nr)}=\tilde{n}\xi^{max}\simeq \max_{\varphi<0}\left[\mu a_0^4(\varphi)+\dot{a}_0(\varphi)\right]
=\left\{\begin{array}{cc}\mu a_0^4,&\mu a_0^3 T\gg 1,\\0.858\, a_0/T, &\mu a_0^3 T\ll 1.\end{array}\right.
\end{equation}
For a gaussian envelop $a_0\propto e^{-\varphi^2/T^2}$ used in this paper the final maximization cannot be done analytically, but only numerically, because of occurrence of a transcendental equation. The limiting cases above are also given for this particular case. Note that in the nonrelativistic case RF becomes dominant if 
\begin{equation}\label{rfeffnr}
\mu a_0^3T\gtrsim 1,
\end{equation}
i.e. for intensities $I\gtrsim 10^{22}$ W/cm${}^2$ if we assume the FWHM pulse duration $150$ fs  as announced for ELI Beamlines \cite{ELI}. 

\subsubsection{Ultrarelativistic electrons: RF induced charge separation.}

From now on assume that longitudinal electron motion is ultrarelativistic, $\gamma\approx u_x\gg a_0$. Then we have: $1-v_x\approx a_0^2/2u_x^2$, $1+v_x\approx2$, $\xi\approx \tau$, so that (\ref{eqcp}) reduces to 
\begin{equation}\label{eqcp_ur}
\frac{du_x}{d\tau}=\frac1{2 u_x}\frac{d(a_0^2)}{d\varphi}+\frac{\mu a_0^6}{4u_x^2}-\tilde{n}\tau.
\end{equation}

Let us consider first the RF mechanism (RFM) of charge separation and neglect the first (ponderomotive) term in the RHS of (\ref{eqcp_ur}). The process obviously splits into two stages. During the {\it acceleration stage} $0<\tau<\tau_{acc}$ the last electrostatic term in the RHS remains small and can be also neglected, hence $du_x/d\tau\approx \mu a_0^6/4u_x^2$, or $u_x\sim a_0^2(\mu\tau)^{1/3}$. The acceleration stage ends up at about the moment 
\begin{equation}\label{tacc_rf}
\tau_{acc}^{(RFM)}\sim(\mu a_0^6/\tilde{n}^3)^{1/5},
\end{equation} 
when the electrostatic term grows to the same order as the RF term. Next, during the {\it deceleration stage} $\tau_{acc}<\tau<\tau_{bd}$,  electron steadily decelerates until the {\it breakdown} at $\tau=\tau_{bd}$, when it finally leaves the laser pulse and rapidly accelerates backward by the electrostatic field. For the deceleration stage by neglecting the LHS in (\ref{eqcp_ur}) we obtain $u_x\simeq\sqrt{\mu/\tilde{n}\tau}a_0^3/2$. The breakdown time can be estimated from the condition that the total phase acquired by the electron in the pulse equals the pulse duration: 
\begin{equation}\label{tbd_cond}
\int\limits_0^{\tau_{bd}}(1-v_x)d\tau=T.
\end{equation}
Since for the laser and plasma parameters providing an efficient RF induced charge separation we have $\tau_{acc}\ll \tau_{bd}$ (see below), in our estimation we can neglect the duration of the acceleration stage and obtain (see \cite{arxiv})
\begin{equation}\label{tau_bd_rfm}
\tau_{bd}^{(RFM)}\simeq\sqrt{\frac{\mu T}{\tilde{n}}}a_0^2.
\end{equation}
Equation (\ref{tau_bd_rfm}), together with
\begin{equation}\label{arfm}
a_\parallel^{(RFM)}=\tilde{n}\tau_{bd}^{(RFM)}\simeq \sqrt{\mu\tilde{n}T}a_0^2,
\end{equation}
define the period and the amplitude of a longitudinal wave generated in a plasma.  

\subsubsection{Ultrarelativistic electrons: ponderomotive charge separation.}

Now let us assume instead that charges are separated by the ponderomotive mechanism (PM) and neglect the second term in the RHS of (\ref{dux}). Estimating the derivative of the envelop \footnote{Here for simplicity we drop all the numerical factors.} by $d a_0^2/d\varphi\sim a_0^2/T$, we can estimate $u_x\sim a_0\sqrt{\tau/T}$ for the acceleration stage and $u_x\sim a_0^2/(\tilde{n}\tau T)$ for the deceleration stage. The acceleration and breakdown times can be estimated in the same manner as above: 
\begin{equation}\label{taupm}
 \tau_{bd}^{(PM)}\sim\tau_{acc}^{(PM)}\sim \left(\frac{a_0^{2}}{\tilde{n}^{2}T}\right)^{1/3}.
\end{equation}
Since in this case the acceleration and breakdown times are of the same order, we can use $\tau_{bd}^{(PM)}$ as a rough order-of-magnitude estimation for the period of a longitudinal wave. The amplitude of the longitudinal field is then given by (see also \cite{arxiv})
\begin{equation}\label{apm}
a_\parallel^{(PM)}\simeq \tilde{n} \tau_{bd}^{(PM)}\simeq\left(\frac{a_0^{2}\tilde{n}}{T}\right)^{1/3}.
\end{equation}

Comparing (\ref{arfm}) and (\ref{apm}) we conclude that in a transparent plasma longitudinal wave generation is enhanced by RF (RFM dominates over PM) if 
\begin{equation}\label{rfeff}
\mu^3\tilde{n}T^5a_0^8\gtrsim 1,
\end{equation}
i.e. for stronger and longer laser pulses. 

\subsection{Linear polarization}
Assuming that radiation damping is small, for a linearly polarized laser pulse $\mathbf{A}(\varphi)=(m/e)a_0(\varphi)\{0,\cos\varphi,0\}$ we have $\mathbf{u}\approx\{u_x,a_0(\varphi)\sin\varphi,0\}$. Hence the equation for longitudinal motion (\ref{dux}) in ultrarelativistic case takes the form 
\begin{equation}\label{eq_lp}
\frac{du_x}{d\tau}\approx\frac{1}{2u_x}\frac{d a_0^2}{d\varphi}\sin^2\varphi+\frac{a_0^2}{2u_x}\sin 2\varphi+\frac{\mu a_0^6\sin^4\varphi\cos^2\varphi}{4u_x^2}-\tilde{n}\tau.
\end{equation}
Unlike the case of circular polarization, the motion is accompanied by fast oscillations, but their envelope behaves qualitatively very similar to that case. During the deceleration stage an approximate solution to (\ref{eq_lp}) can be represented in the form\footnote{Unlike the case of circular polarization, we can neglect in this stage the derivative of the envelop $C$, but not of the remaining rapidly oscillating factor.} $u_x\simeq C(\tau)\sin^2\varphi$. As we will shortly see, for $|\varphi|\sim T\gg 1$ we have $\tau\dot{\varphi}\simeq \varphi\gg 1$, indicating that $\tau$ (and hence $C$) is a slowly varying function of $\varphi$. Hence to find $C(\tau)$ we can average (\ref{eq_lp}) over the laser oscillations. The results are  $C(\tau)=a_0^3\sqrt{\mu/(8\tilde{n}\tau)}$ for RFM and $C(\tau)=a_0^2/(2\tilde{n}T\tau)$ for PM. Now we can calculate $\tau(\varphi)$ for both  cases from 
\begin{equation}\label{dtau}
\frac{d\tau}{d\varphi}\approx \left\langle\frac{2u_x^2}{(a_0\sin\varphi)^2}\right\rangle=\frac{C^2(\tau)}{a_0^2},
\end{equation} 
thus obtaining $\tau(\varphi)\propto \varphi^{1/2}$ for RFM and $\tau(\varphi)\propto \varphi^{1/3}$ for PM. Power dependence of $\tau$ on $\varphi$ validates the assumption of slow variation of $\tau$ for $|\varphi|\sim T\gg1$. From $\tau_{bd}=\tau(T)$ we obtain that after the substitution $a_0\to a_0/\sqrt{2}$ equations (\ref{tau_bd_rfm})--(\ref{apm}) remain valid for linear polarization as well. Hence in a dilute plasma the charge separation field strength, expressed in terms of laser intensity, 
\begin{equation}
a_\parallel^{(RFM)}\simeq\sqrt{\mu\tilde{n}T}\frac{I}{I_0},\quad a_\parallel^{(PM)}=\left(\frac{\tilde{n}}{T}\frac{I}{I_0}\right)^{1/3},
\end{equation}
where $I_0=\pi m^2c^5/(e^2\lambda^2)\approx 2.74\cdot 10^{18}$ W/cm${}^2$ (for $\lambda=1\mu$m), is independent of the laser polarization. 

\subsection{Analysis of used approximations}\label{sec_ca}

\begin{figure}[t!]
\subfloat{\includegraphics[width =0.5\linewidth]{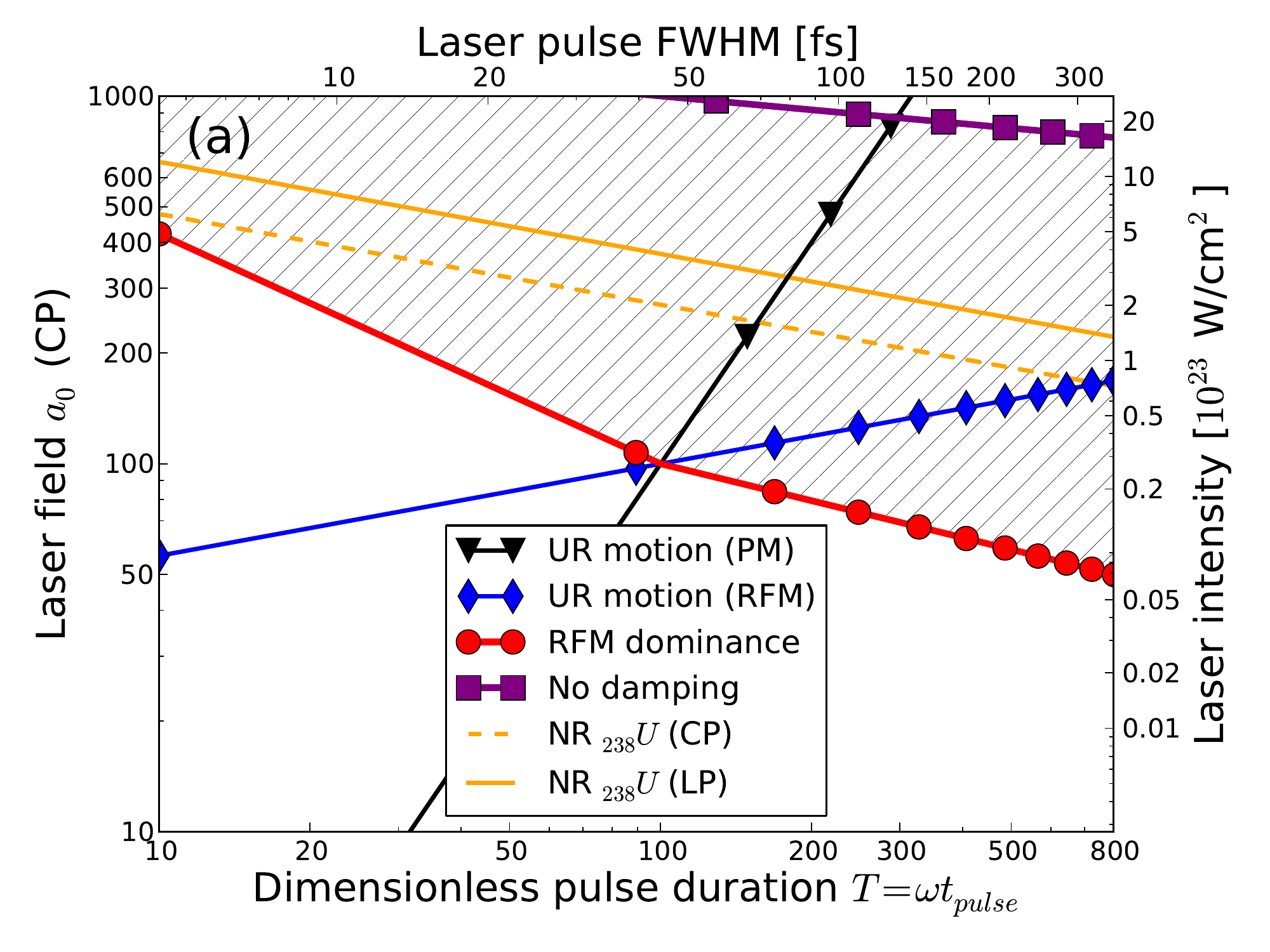}}
\subfloat{\includegraphics[width =0.5\linewidth]{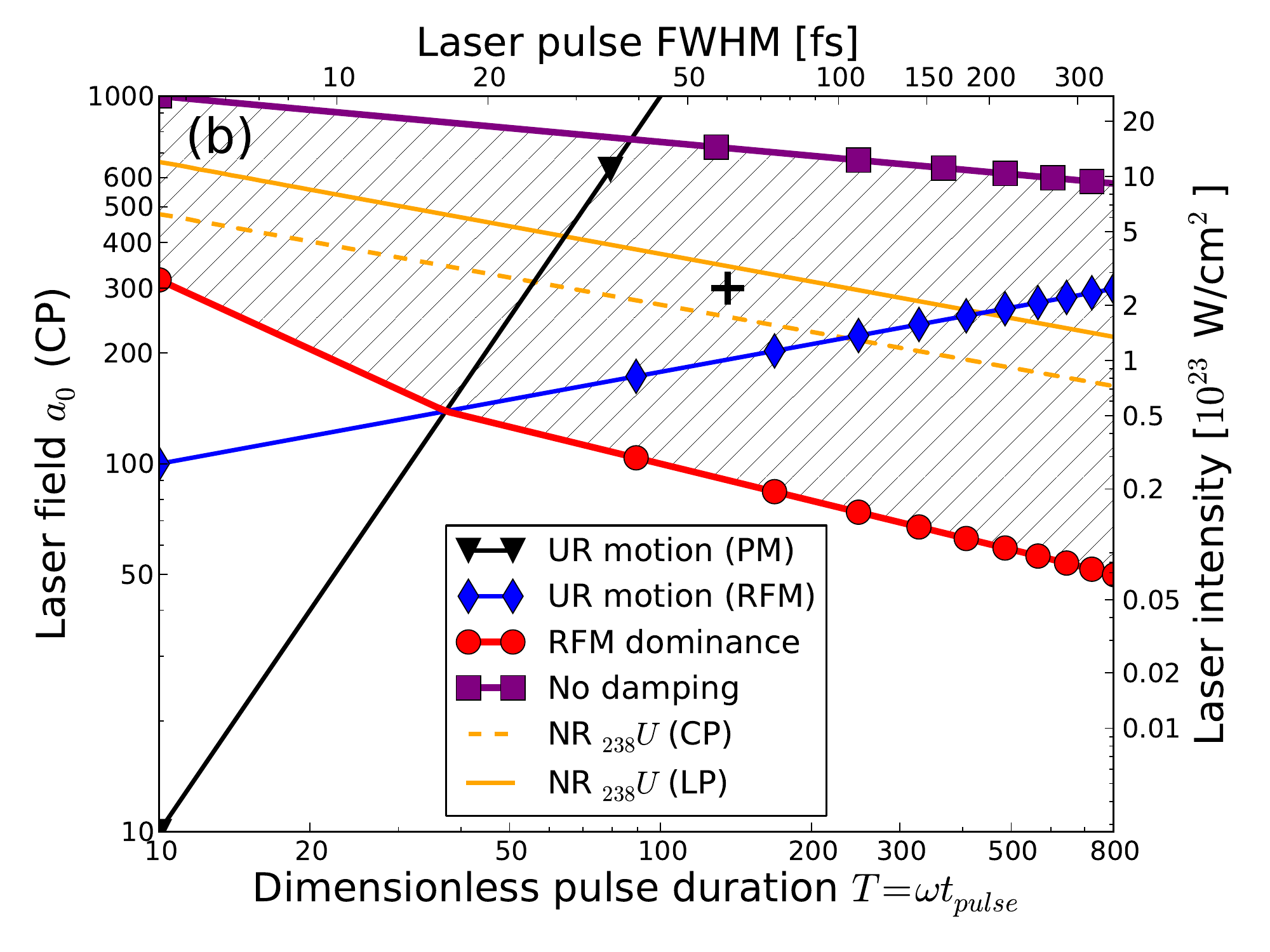}}
\caption{\label{fig2}(Color online).  A map of various regimes of the longitudinal field generation with respect to laser pulse intensity and duration according to Sec.~\ref{sec_ca} for plasma densities: (a) $n=0.01 n_c$ and (b) $n=0.1 n_c$. Markers on the lines: black triangles and blue diamonds bound from below ultrarelativistic (UR) electron motion for PM and RFM, respectively [see (\ref{ur})]; purple squares bound from above the region of low damping of the laser pulse [see (\ref{damping})]; red circles bound from below the region of RFM dominance over PM [see (\ref{rfeffnr}) and (\ref{rfeff})]. The yellow unmarked lines
bound from above the region of nonrelativistic (NR) ${}_{238}U$ ions according to (\ref{nrions}) with respect to the (right) intensity axis for the cases of linear (upper solid line) and circular (lower dashed line) polarization.}
\end{figure}

Let us discuss in more detail and summarize the conditions of validity of the approximations we made while deriving the estimations (\ref{arfm}) and (\ref{apm}): 
\begin{itemize}
\item The laser field is strong, $a_0\gg1$.
\item Longitudinal motion is ultrarelativistic, $u_x(\tau_{bd})\gg a_0$, or 
\begin{equation}\label{ur}
\left(\frac{\mu}{\tilde{n}T}\right)^{1/4}a_0\gg1,\quad \mathrm{or}\quad \left(\frac{a_0}{\tilde{n}T^2}\right)^{1/3}\gg1\
\end{equation}
for RFM and PM cases, respectively [as they should, these conditions are opposite to the ones in (\ref{nrcond})]. Note that while the period of the created quasistatic longitudinal wave $\simeq T$ in the regime of nonrelativistic longitudinal electron motion, under the conditions (\ref{ur}) according to (\ref{tau_bd_rfm}) and (\ref{taupm}) it is $\gg T$.
\item Transverse radiation damping is small, i.e. the dimensionless transverse RF force $F_{RF}\simeq\mu a_0^2\gamma^2(1-v_x)v_\bot$ is much smaller than the dimensionless transverse Lorenz force $F_L=a_0$. In the ultrarelativistic case it is equivalent to $\mu a_0^4/u_x\ll1$ or, taking into account that $u_x\gtrsim u_x(\tau_{bd}^{(RFM)})$, to
\begin{equation}\label{damping}
(\mu^3\tilde{n} T)^{1/4} a_0^2\ll1.
\end{equation}
\item For RFM, the acceleration time is much smaller than the breakdown time. In fact, this assumption turns out to be literally equivalent to the condition (\ref{rfeff}) of dominance of RFM over PM, see (\ref{tacc_rf}) and (\ref{tau_bd_rfm}). 
\item Depletion of the transverse laser field is neglected, i.e. we assume that the (dimensionless) energies $\varepsilon_e$ gained by plasma electrons and $\varepsilon_\parallel$ accumulated in the quasistatic longitudinal field are much smaller than the (dimensionless) energy of the laser pulse $\varepsilon_L\simeq a_0^2T$. The number of accelerated electrons can be estimated as $\tilde{n}\tau_{bd}$ and the (dimensionless) average energy of a single electron as $u_x(\tau_{bd})$, hence for RFM we have $\varepsilon_e\simeq (\mu^3\tilde{n}T)^{1/4}a_0^4$ and $\varepsilon_\parallel\simeq a_\parallel^2\tau_{bd}\simeq \sqrt{\mu^3T^3\tilde{n}}a_0^6$. Note that: 
\subitem if the condition (\ref{rfeff}) of RFM dominance is fulfilled, then $\varepsilon_\parallel\gtrsim \varepsilon_e$, meaning that RFM is an efficient mechanism for longitudinal field generation; 
\subitem the condition (\ref{damping}) of weak damping is equivalent to $\varepsilon_L\gg\varepsilon_\parallel$ of negligibility of the laser field depletion. 

\item Immobility of the ions. In reality, the generated longitudinal electric field not only decelerates the electrons, but also accelerates the ions. We can neglect the ion motion in the estimations (\ref{tau_bd_rfm})--(\ref{apm}) only if ions remain nonrelativistic until the electron breakdown. In such a case the equation of ion motion is of the form $\ddot{\xi}_i(\tau)=\xi_i(\tau)/\tau_i^2$ and its solution reads $\xi_i(\tau)\simeq \xi_0 e^{\tau/\tau_i}$, where 
$$ \tau_i=\sqrt{\frac{Am_p}{Z_*m_e\tilde{n}}},$$ 
$\xi_0$ is the initial position of an ion, $m_p$ is the nucleon mass, $A$ and $Z_*e$ are the ionic weight and charge. An estimation of the ionic charge which we used in this paper is discussed in the Appendix. Since ion acceleration is initially exponential, the ions become relativistic within the time of order $\tau_i$. Hence the condition that ions remain nonrelativistic (and can be considered immobile) is written as $\tau_i\gg \tau_{bd}$ or, explicitly,
\begin{equation}\label{nrions}
a_0\ll\left(\frac{m_p}{m_e}\frac{A}{Z_*}\frac{1}{\mu T}\right)^{1/4}.
\end{equation}
When the ions become relativistic the charge separation and the longitudinal field should get saturated. 
\end{itemize}

The resulting domains of validity of all the independent assumptions discussed above for two plasma densities $n=0.01 n_c$ and $n=0.1 n_c$ are illustrated in Figure~\ref{fig2} together with the line separating the dominance of RFM over PM. For clarity, the target region between this line and the topmost line of strong pulse depletion is shaded. The values of intensities from the right axis are converted to the equivalent values of $a_0$ for circular polarization (CP) on the left axis. One can see that for relatively long pulses ($t_{pulse}\gtrsim 100$ fs) RF is dominant already at intensities exceeding $10^{22}$ W/cm${}^2$. Since ionization degree depends on the field strength but not on the intensity, the bottom orange line without markers, separating the relativistic and nonrelativistic regimes of ion motion for circular polarization, is drawn with respect to both left and right axis, but for linear polarization with respect to the right axis should be replaced with the top one. 

\subsection{Optimal density and maximally attainable charge separation field}

According to (\ref{arfm}) and (\ref{apm}), for a dilute plasma and given laser parameters $a_0$ and $T$, the charge separation field is growing monotonically with the plasma density. However, in virtue of (\ref{nrcond}) and/or (\ref{ur}), for 
\begin{equation}\label{nopt}
\tilde{n}\gtrsim\tilde{n}_{opt}\simeq \mu a_0^4/T
\end{equation}
the electron longitudinal motion becomes nonrelativistic and hence for higher densities the charge separation field is density independent\footnote{Note, however, that for densities (\ref{nopt}) the plasma effects ignored in our model can be important.}, see (\ref{anr}). Even though further increase of plasma density can no more affect the maximal charge separation field, it may enhance the laser pulse damping. Hence the longitudinal charge separation field generated in a plasma is maximal for the optimal density (\ref{nopt}) and is bounded by 
\begin{equation}\label{amax}
a_\parallel\lesssim a_\parallel^{(max)}\simeq \mu a_0^4\ll a_0,
\end{equation}
where we used that in the case of optimal density (\ref{nopt}) the condition of negligible damping (\ref{damping}) reduces to $\mu a_0^3\ll 1$. The upper limit $a_\parallel\lesssim a_0$ could be probably approached when damping is moderate $\varepsilon_\parallel\lesssim \varepsilon_L$, i.e. beyond the applicability of our current model.

\section{PIC simulations results}

In order to confirm our theoretical findings we performed one and two dimensional PIC simulations using the code EPOCH \cite{EPOCH} with the classical radiation friction included in the form of Landau and Lifshitz \cite{arxiv}. The numerical approach is similar to the one developed in \cite{zhidkov2002,tamburini2010}. 

\begin{figure}[th!]
\centering
\includegraphics[width =0.6\linewidth]{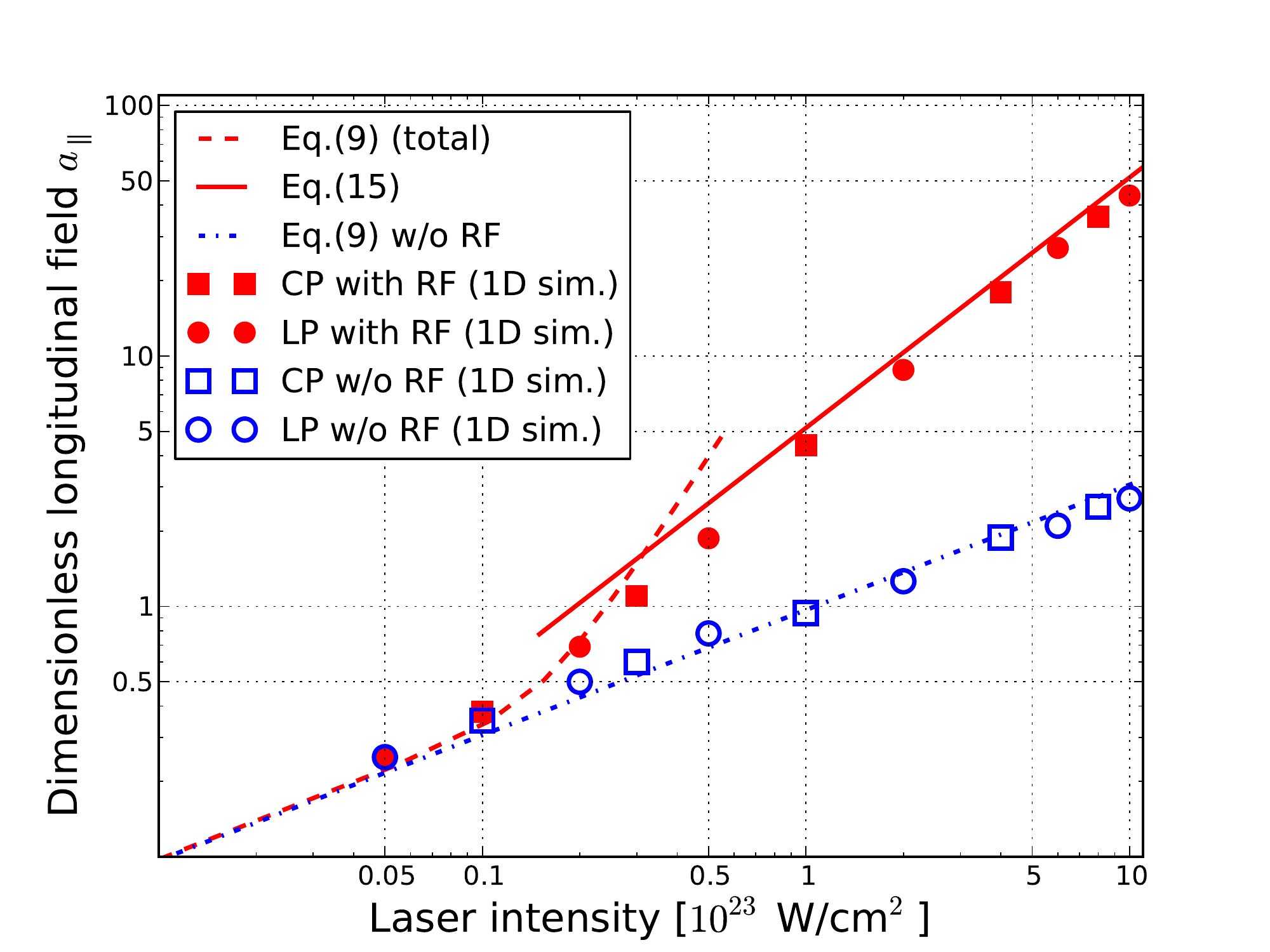}
\caption{\label{fig3}(Color online). Comparison of our theoretical predictions for an amplitude of the generated longitudinal field with the 1D PIC simulation data [with (red filled markers) and without (blue empty markers) RF]. The markers of square and circle shapes correspond to the results of simulation with circularly and linearly polarized driving pulses of the given peak intensity, respectively. The red solid line -- estimation (\ref{arfm}); red dashed line -- approximation (\ref{anr}); blue dash dotted line corresponds to the first term from the RHS in (\ref{anr}). FWHM pulse duration $150$ fs, plasma density $n=0.01 n_c$, immobile ions.}
\end{figure}

The results for the dependence of the longitudinal field amplitude on laser intensity from 1D simulations are presented in Figure~\ref{fig3}. We used 100 cells per wavelength and 20 particles per cell. The laser pulse envelop was chosen Gaussian with FWHM duration $150$ fs, as expected at ELI Beamlines \cite{ELI}. The initial plasma density was $n=0.01 n_c$ and the ions were considered immobile. As one can observe, when expressed in terms of intensity, the longitudinal field generation is insensitive to laser polarization, and that the simulation data is in excellent agreement with the approximation (\ref{anr}) and the estimation (\ref{arfm}). At the same time, neglecting RF results in substantial (up to an order of magnitude) underestimation of the longitudinal field already for the intensities $I\gtrsim 2\cdot 10^{22}$ W/cm${}^2$. Note that for such long pulses electron acceleration in longitudinal direction when RF is neglected (i.e. solely due to PM) remains nonrelativistic for the whole shown laser intensity range. However, by considering shorter pulses we have previously justified in \cite{arxiv} the estimation (\ref{apm}) for PM for ultrarelativistic case as well.

\begin{figure}[t!]
\subfloat{\includegraphics[width = 0.5\linewidth]{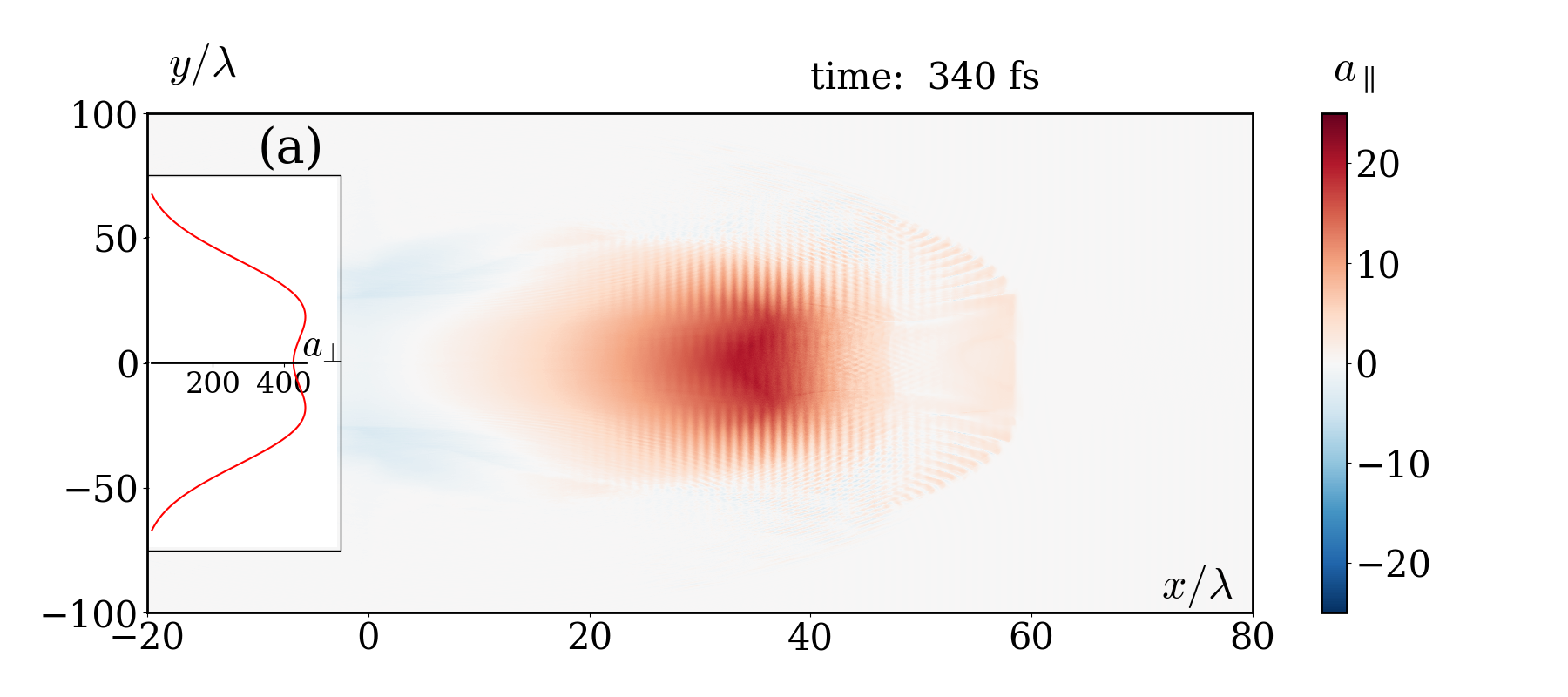}} 
\subfloat{\includegraphics[width = 0.5\linewidth]{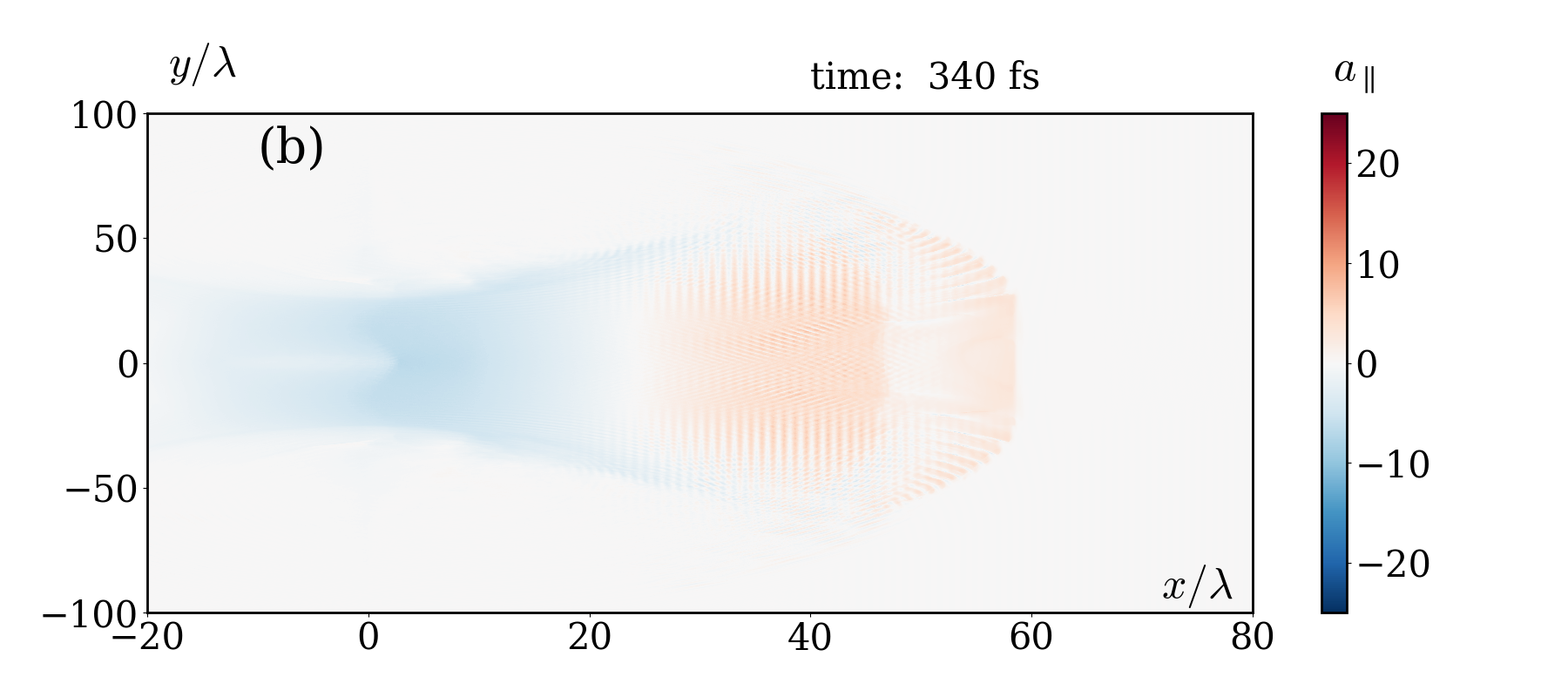}}\\
\subfloat{\includegraphics[width = 0.5\linewidth]{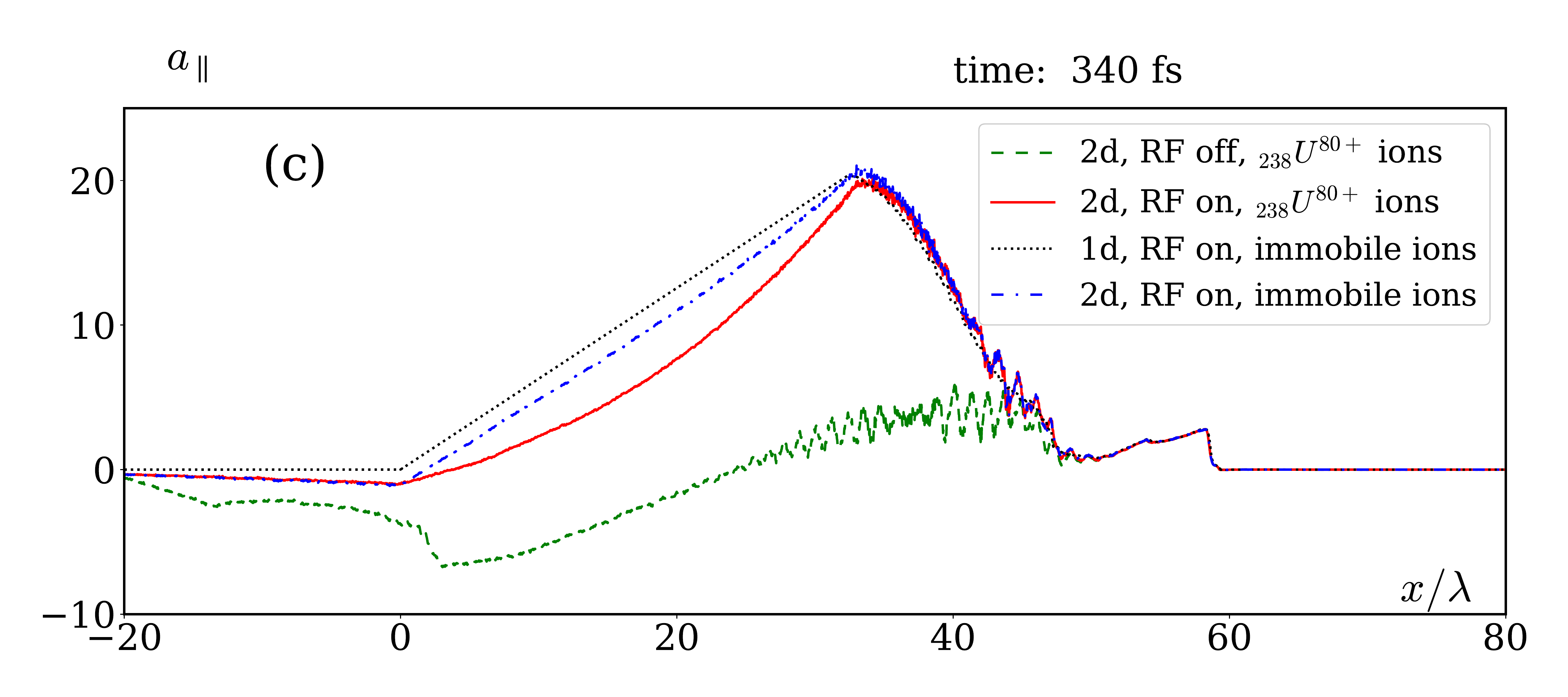}}
\subfloat{\includegraphics[width = 0.5\linewidth]{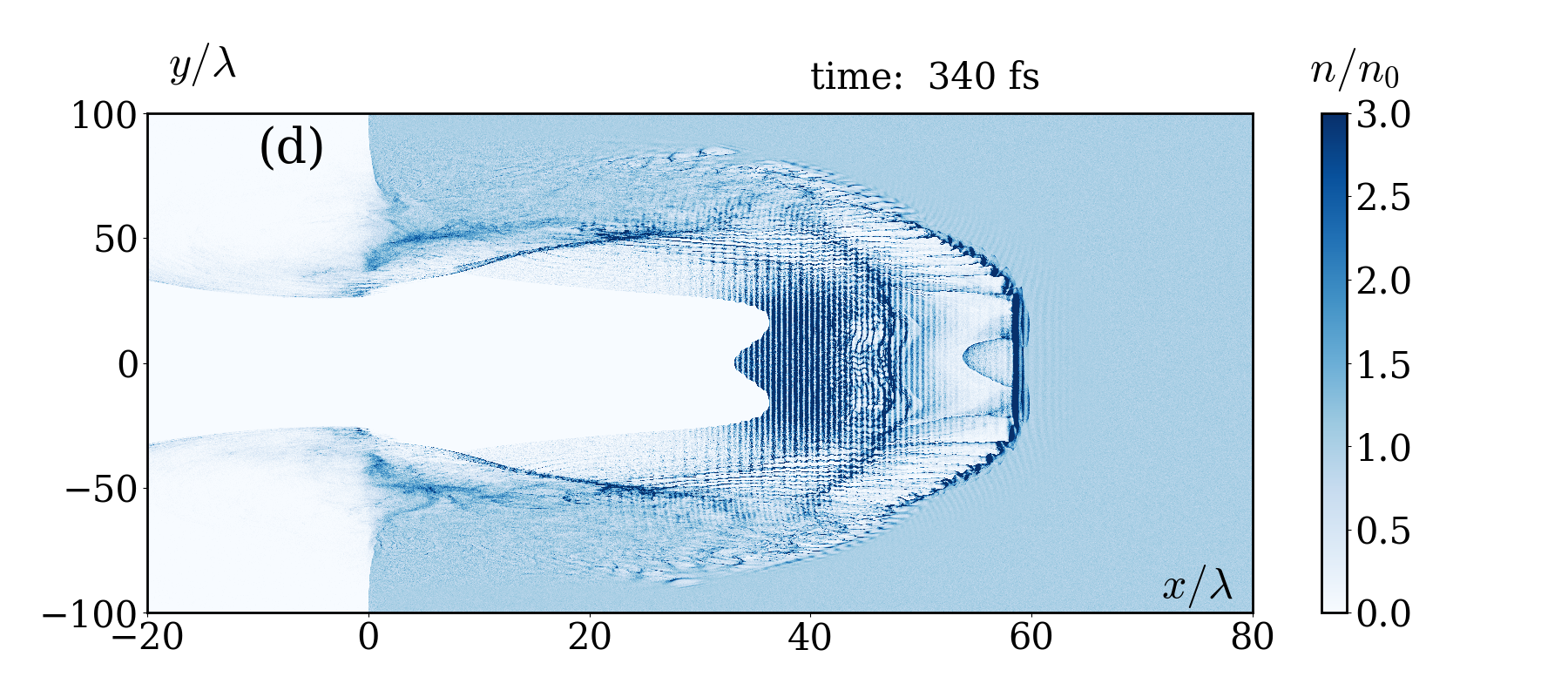}}
\caption{\label{fig4}(Color online). Snapshot of 2D PIC simulation of a wide laser pulse propagating in a plasma taken at the moment of breakdown $t=340$ fs. (a) and (b) -- longitudinal field distributions with and without RF (inset -- laser pulse transverse profile); (c) -- longitudinal electric field on the $x$-axis from: 1D and 2D PIC simulations with account for RF and with immobile ions (black dotted curve and blue dash dotted curve, respectively), 2D PIC simulations with and without RF and with mobile ${}_{238}^{92}U^{80+}$ ions (red solid curve and green dashed curve, respectively); (d) -- electron density distribution with account for RF. Laser intensity $I=2.5\cdot10^{23}$ W/cm$^2$, FWHM $60$ fs, linear polarization, $n=0.1 n_c$, other parameters specified in the text.}
\end{figure}

In order to strengthen the reliability of our results, we performed also a series of more realistic 2D EPOCH simulations\footnote{For 2D PIC simulations we use 50 cells per wavelength and 10 particles per cell.} with linearly polarized driving pulses (for 2D simulations with circularly polarized pulses see \cite{arxiv}) and the heavy mobile ions $_{238}^{92}U^{80+}$ (the average ionization degree is estimated according to the Appendix). As in \cite{arxiv}, in order to reduce transverse expulsion of electrons by the ponderomotive force of the pulse, in all simulations we use the pulses with a symmetric bimodal Gaussian transverse profile shown in the inset of the Figure~\ref{fig4}~(a), with the distance $d$ between the peaks fixed so that the transverse field amplitude on the $x$-axis coincides with the amplitude of a single pulse, $d=2w\sqrt{\ln{2}}\approx 1.7w$, where $w$ is the waist of each peak.

Our first simulation (see Figure~\ref{fig4}) corresponds to the case of a wide (weakly focused of comparable width and length, $w=t_{pulse}=22\lambda$, where the wavelength $\lambda=1\mu$m) driving pulse. Since in this particular case we are mainly interested in stronger validation of our 1D model, we determined the simulation parameters according to Figure~\ref{fig2} to maintain the longitudinal electron motion ultrarelativistic, while keeping ions nonrelativistic: plasma density $n=0.1 n_c$, laser intensity $I=2.5\cdot10^{23}$ W/cm${}^2$ and FWHM $60$ fs -- see the cross between the unmarked orange solid line and the blue line marked with diamonds in the Figure~\ref{fig2}~(b).  \label{param_fig4}  From Figures~\ref{fig4}~(a) and (b), where we compare the longitudinal fields computed with and without accounting for RF, respectively, one can observe that for these parameters the effect of RF-induced enhancement is extremely well pronounced in 2D. Moreover, for such wide pulses the longitudinal field distribution on the $x$-axis is in a rather good agreement with both the 1D simulations and the estimations of the preceding section, see Figure~\ref{fig4}~(c). The most notable 2D effect is that some electrons bypass the ion bubble \cite{bubbles}, getting inside from its rear side [see Figure~\ref{fig4}~(d)], and in this way screening the quasistatic longitudinal field. Its slight decrease (as compared to the 1D simulation) at the rear side of the resulting longitudinal wave in Figure~\ref{fig4}~(c) is explained partially by this effect, and partially by the nonrelativistic ion motion. 

\begin{figure}[t!]
\subfloat{\includegraphics[width = 0.5\linewidth]{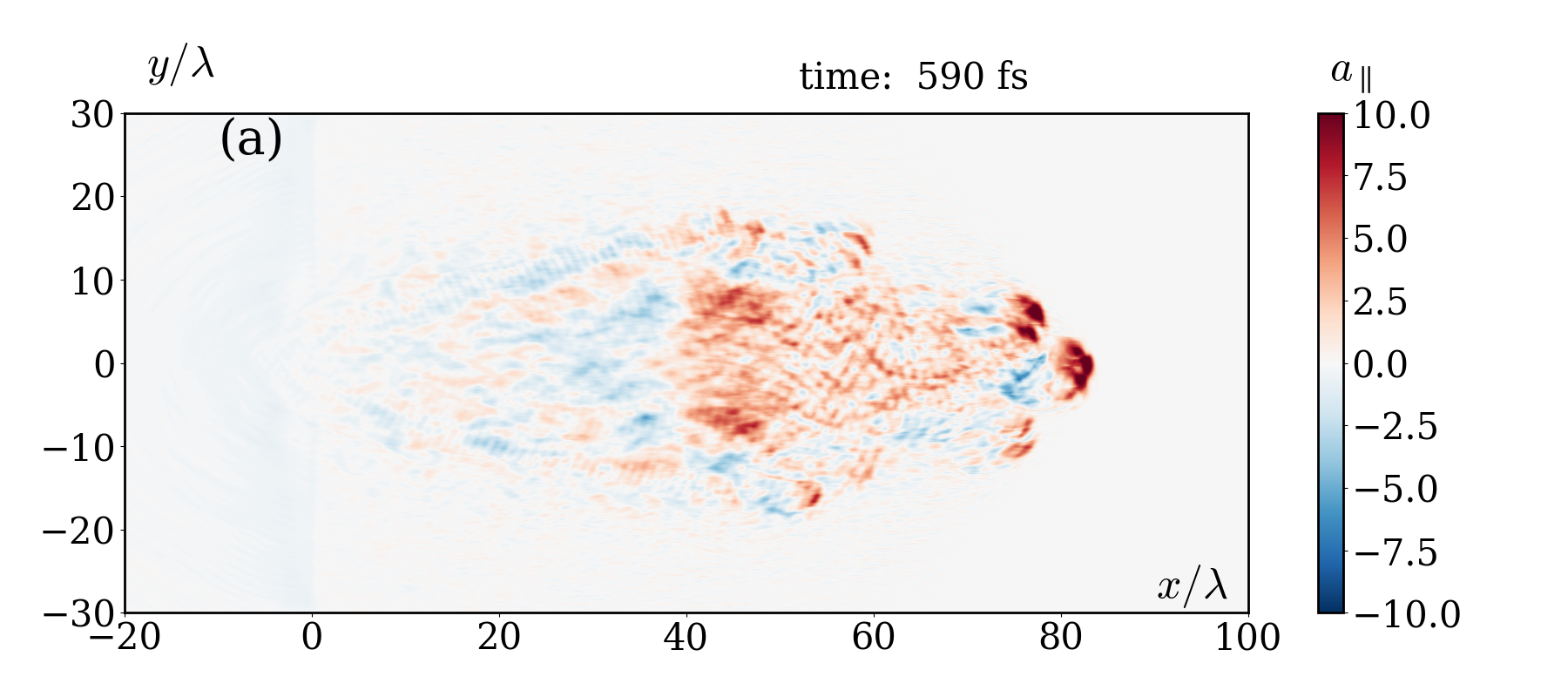}} 
\subfloat{\includegraphics[width = 0.5\linewidth]{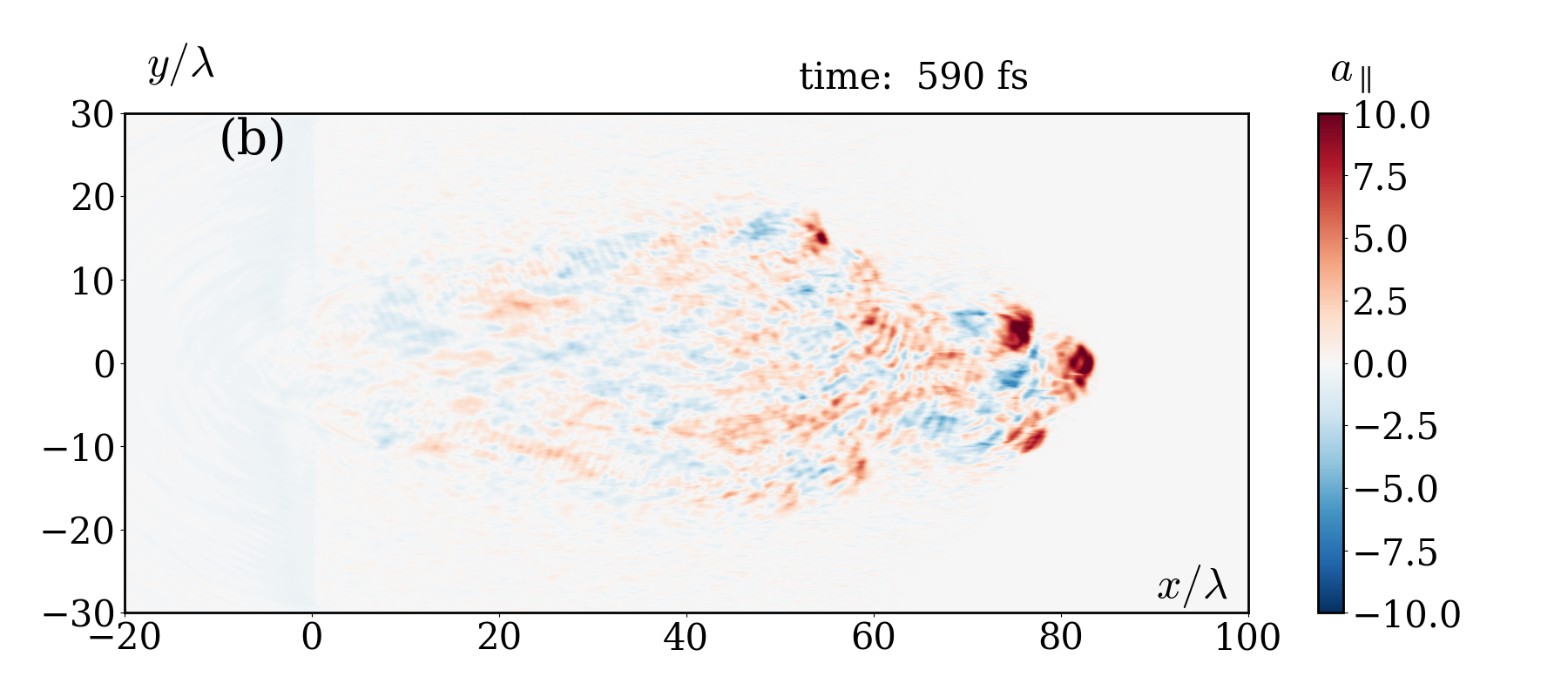}}\\
\subfloat{\includegraphics[width = 0.5\linewidth]{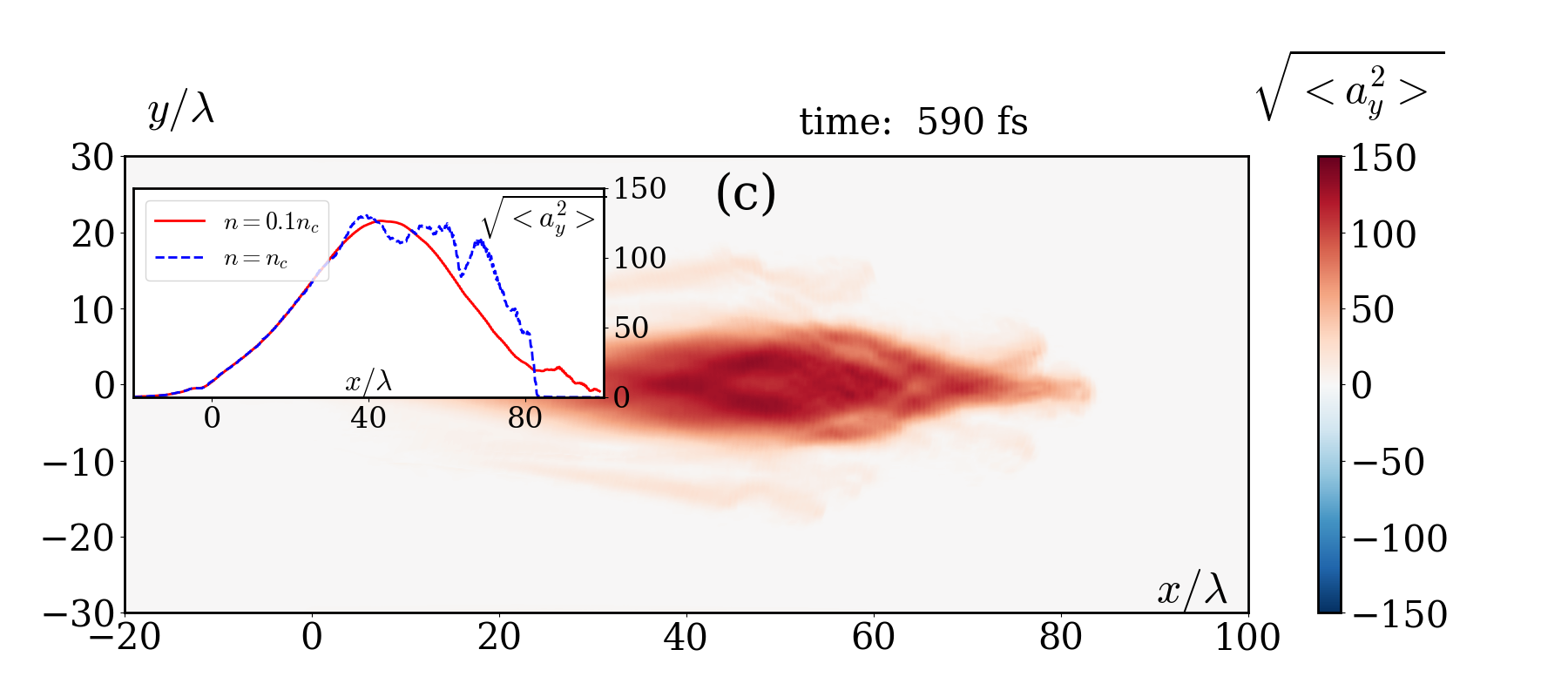}}
\subfloat{\includegraphics[width = 0.5\linewidth]{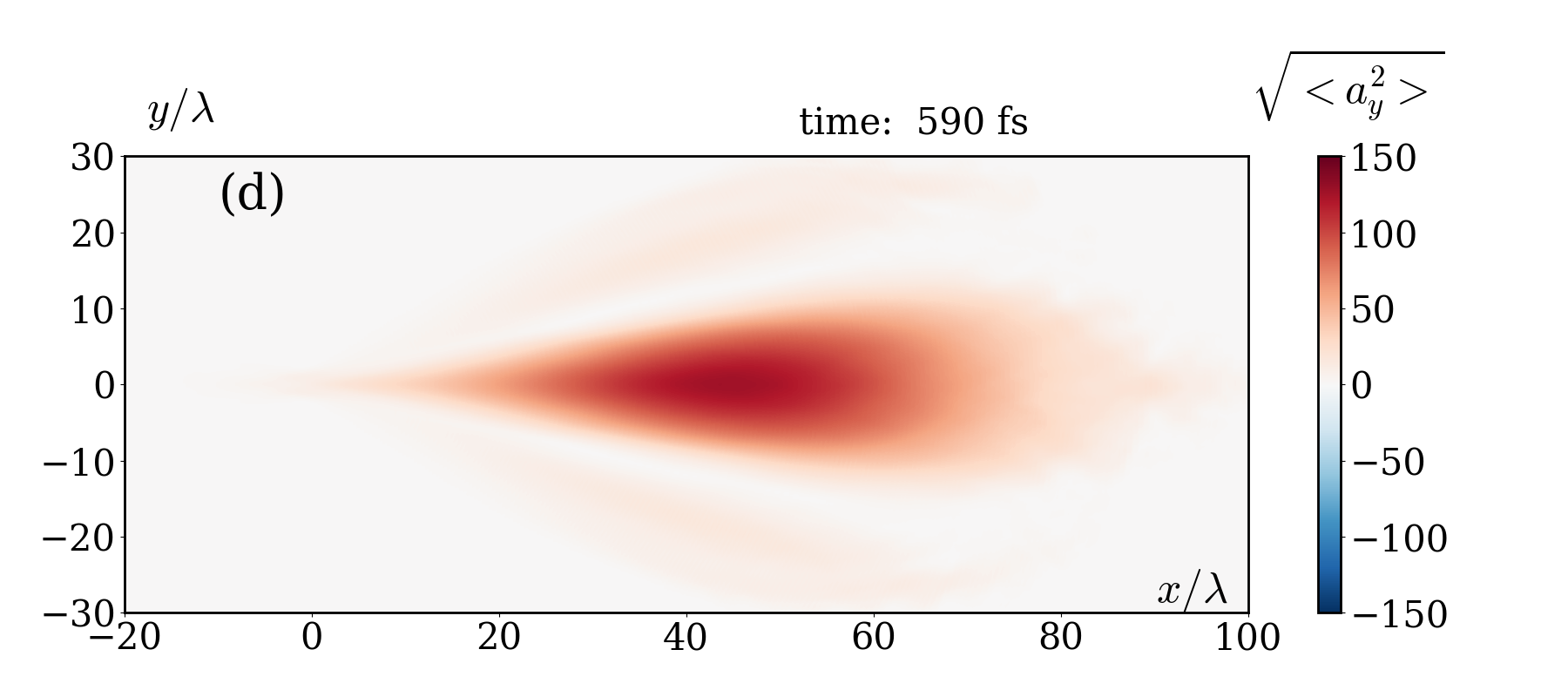}}
\caption{\label{fig5}(Color online). Snapshot of 2D PIC simulation of a tightly focused ($w=\lambda=1\mu$m) laser pulse propagating in a plasma with mobile ions $_{238}^{92}U^{80+}$ taken at $t=590$ fs. (a) and (b): longitudinal field distributions, averaged over carrier wavelength along the $x$-axis, with and without RF taken into account, respectively, for plasma density $n=n_c$; (c) and (d): RMS over carrier wavelength of the transverse field for $n=n_c$ and $n=0.1 n_c$, respectively [inset:  RMS of the transverse field on the $x$-axis for $n=n_c$ (blue dashed line) and $n=0.1 n_c$ (red solid line)]. Laser intensity $I=1.7\cdot 10^{23}$ W/cm${}^2$, FWHM $150$~fs, other parameters specified in the text.}
\end{figure}

Since the laser pulses discussed in Figure~\ref{fig4} are both intense and wide, their total power is extremely high.  In order to substantiate the effect in a more realistic setup, we also present the results of simulations with the pulses tightly focused (waist radius $w\simeq\lambda$) at the left plasma boundary, see Figure~\ref{fig5}. Here we assume the parameters announced for ELI Beamlines \cite{ELI}\label{param_tight}: $I=1.7\cdot 10^{23}$ W/cm${}^2$ (total power $\simeq 10$ PW) and FWHM $150$ fs. In order to suppress strong diffraction by selffocusing, keeping longer the field strong enough for a noticeable effect of RF, we choose here plasma density $n=n_c$. The resulting longitudinal fields averaged over laser wavelength -- with and without RF -- are shown in Figures~\ref{fig5}~(a) and (b) [comparison of a root mean square (RMS) of the transverse field distribution with the case $n=0.1n_c$ is shown in Figures~\ref{fig5}~(c) and (d)]. According to the figures, the effect of RF induced enhancement of the longitudinal field generation in a plasma could be observed on the forthcoming 10 PW laser facilities of the near future. 

\section{Discussion}

A laser pulse propagating in plasma separates the charges thus generating a quasistatic longitudinal electric field, which, as discussed in this paper, under certain conditions can be substantially enhanced due to modification of the electron transverse oscillations by even weak radiation friction. To describe the effect, we develop a 1D model and propose explicit analytical estimates of an amplitude and a period for the enhanced longitudinal field according to a particular regime of the process (nonrelativistic versus ultrarelativistic longitudinal electron motion, and RFM versus PM dominance). In particular we show that, being expressed in terms of the laser intensity, these parameters are in fact independent of polarization of the driving pulse. Our theoretical analysis is limited by the assumptions that the damping of the driving laser pulse is small and that ions are immobile, and we provide a detailed analysis of their validity for a wide range of laser and plasma parameters. Due to energy conservation, the amplitude of the generated longitudinal field can never exceed the amplitude of a driving laser field and, according to our theory, for given pulse intensity and duration, the highest longitudinal field is achieved with such an optimal plasma density that the electron longitudinal motion is mildly relativistic. 

We have validated our predictions by comparing them to PIC simulations with both immobile and mobile high-Z ions in 1D, as well as with 2D wide driving pulses. However, consideration of concurrently strong and wide pulses implies unrealistically huge laser power, therefore in order to realize the proposed regime of RF-induced enhancement of the longitudinal field generation within the limitations imposed by the current or foreseeable experimental capabilities, tight focusing is required. For tightly focused pulses our model is no longer valid literally on a quantitative level, because such purely 2D effects as fast diffraction, various plasma instabilities, and the alternating longitudinal selffield of a tightly focused pulse (which can in fact be as strong as the transverse field itself), are vital. Increase of the plasma density, in addition to boosting the magnitude of the generated quasistatic longitudinal field as explained above, also facilitates to suppress pulse diffraction by selffocusing, but is limited by laser radiative damping and by the condition of plasma transparency, as in an opaque plasma electrons cannot penetrate deep inside the pulse to experience a strong enough field for a noticeable RF. According to simulations, by certain tuning of the parameters, the effect of the RF-induced longitudinal field enhancement can be still revealed at the ELI Beamlines facility in the near future \cite{ELI}. 



\ack
We are grateful to M. Grech and M. Vranic for a useful advise on data visualization and to S.V. Popruzhenko for an extremely valuable discussion of the idea underlying our approach for estimation of the ionization degree and for the relevant references. The research was performed using the code EPOCH (developed under the UK EPSRC grants \texttt{\detokenize{EP/G054940/1}}, \texttt{\detokenize{EP/G055165/1}}, and \texttt{\detokenize{EP/G056803/1}}) and the resources of the ELI Beamlines Eclipse cluster, and was partially supported by the MEPhI Academic Excellence Project (Contract No.~\texttt{\detokenize{02.a03.21.0005}}), the Russian Fund for Basic Research (Grants~\texttt{\detokenize{16-32-00863mol_a}} and \texttt{\detokenize{16-02-00963a}}), projects ELITAS (ELI Tools for Advanced Simulation) \texttt{\detokenize{CZ.02.1.01/0.0/0.0/16_013/0001793}} and HiFI (High-Field Initiative) \texttt{\detokenize{CZ.02.1.01/0.0/0.0/15003/0000449}} from European Regional Development Fund.


\begin{appendix}
\section{Estimation of ionization degree}

In order to deal with the ion motion, we need first to estimate the ionization degree (average ionic charges) for a field of a given strength. Even though now EPOCH contains a subroutine for realistic simulation of ionization, for highly-charged ions it generates a large number of species, and correct account for them requires significant computer resources and can seriously slow down simulations. A simplified approach is also needed if we want to make estimations analytically. Hence here we used for this purpose the following heuristic approach.

In a high intensity laser field ionization occurs via the tunneling mechanism, for which general theory is rather well developed (see, e.g. \cite{poprz}). One can use the known formulas for ionization probability $w_i(a_0)$ to estimate the ionization degree from the condition $w_i(a_0)\cdot t_{pulse}\simeq 1$. Since $w_i(a_0)$ depends on the field $a_0$ exponentially, and also because the laser pulse duration is large enough on the atomic scale, this criterion can be roughly reformulated as that ionization of an atomic level takes place when the laser field strength exceeds about $10\%$ of the atomic field strength at the corresponding  outer shell [the actual fraction weakly (logarithmically) depends on the parameters and can be specified more accurately if needed, see below]. 
\begin{figure}[t!]
\centering
\includegraphics[width =0.6\linewidth]{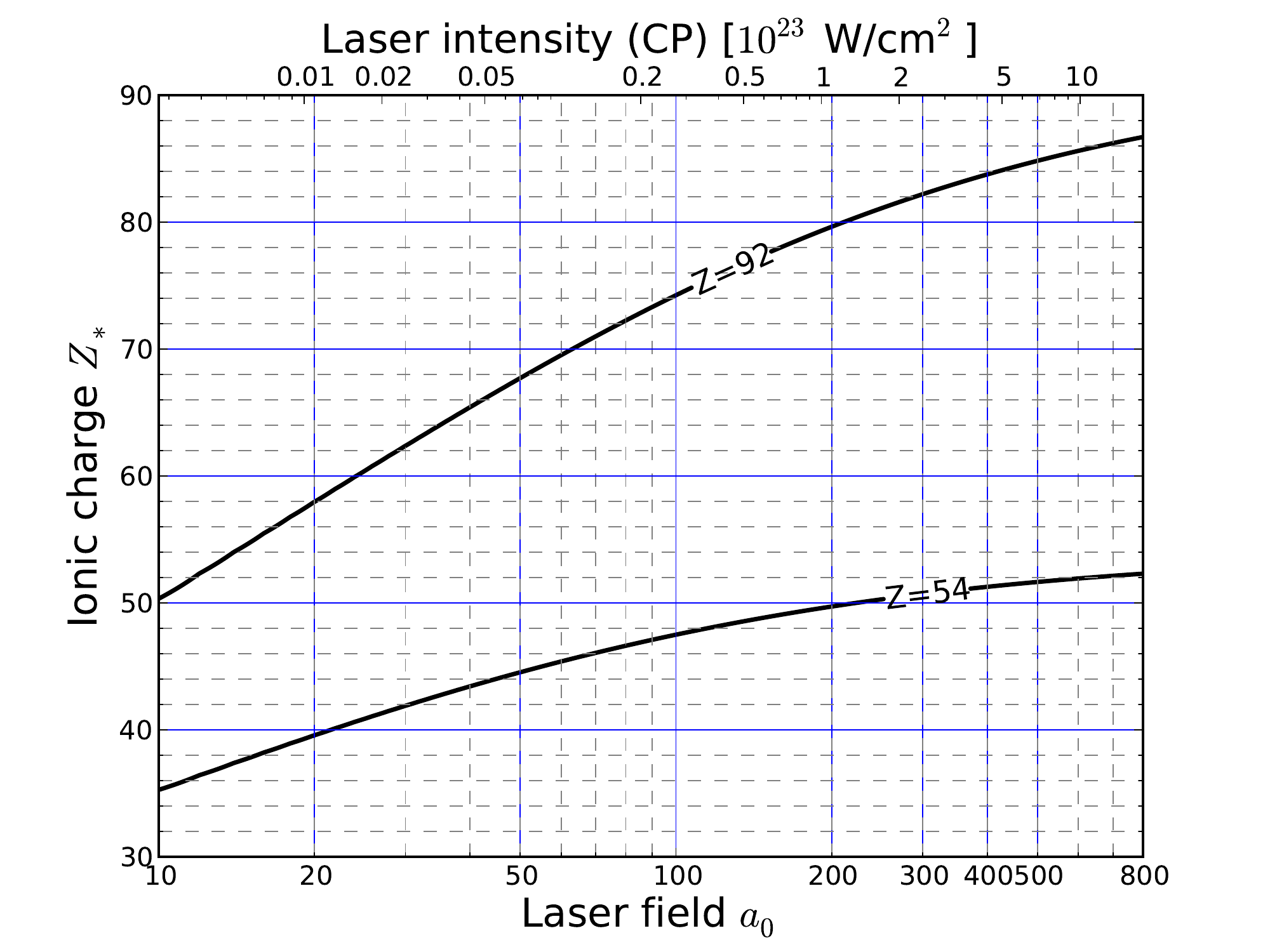}
\caption{\label{fig6}(Color online). Dependence of the average ionic charges of the ions $^{92}_{238}U^{Z_*+}$ and $^{54}_{131}Xe^{Z_*+}$ on the dimensionless laser field strength $a_0$ according to (\ref{i-degree}). The upper axis converts $a_0$ to an equivalent value of intensity for a CP case.}
\end{figure}

Let us consider an atom with atomic number $Z$ and denote its residual ionic charge by $Z_*$, meaning that $Z-Z_*$ electrons remain, while $Z_*$ electrons have escaped due to ionization. We can roughly estimate the principle quantum number $n_{max}$ of the outer shell by using the Pauli principle and the known degeneracies of a Hydrogen-like ion:
$$Z-Z_*=\sum\limits_{n=1}^{n_{max}} 2n^2\simeq \frac23n_{max}^3.$$
Assuming further that the corresponding fraction of the nuclei charge is completely screened by 
residual electrons, we can estimate the outer shell radius as $R_{out}\sim (\hbar^2/m Z_*e^2) n_{max}^2$ and the corresponding atomic field strength as $Z_*e/R_{out}^2=m^2e^5Z_*^3/(\hbar^4n_{max}^4)$.  
With all that our criterion can be formulated as 
\begin{equation}\label{i-degree}
a_0\simeq 0.1\frac{m}{\hbar\omega}\frac{(\alpha Z_*)^3}{\left[\frac32(Z-Z_*)\right]^{4/3}}.
\end{equation}
The solution to (\ref{i-degree}) for Uranium ($Z=92$) and Xenon ($Z=54$) ions that were used in the work on this paper are plotted in Figure~\ref{fig6}. 

A more advanced approach could include: more accurate consideration of the mentioned large logarithmic factor (resulting from the known pre-exponential factor of the tunneling formula), an account for relativistic corrections, and a two-stage procedure of first determining an ionization potential from the tunneling formula, and next selecting the corresponding ion using the tabular data (e.g. from \cite{nist}). We have checked, however, that all these complications do not change our heuristic results within $\simeq 10\%$, which is enough for the rough estimates presented in the paper.
\end{appendix}

\end{document}